\documentclass[aps,graphicx,12pt,showkeys,showpacs]{revtex4}
\usepackage{amsmath}
\usepackage{subfigure}
\usepackage{amscd}
\usepackage{graphicx}

\usepackage{CJK}   
\usepackage{indentfirst}  
\usepackage{bm}    
\usepackage{graphicx}  
\usepackage{epsfig} 
\usepackage{multirow}
\usepackage{makecell}

\begin{document}

\title[Short Title]{One-step direct measurement of the entangled W states with cross-Kerr nonlinearity}

\author{Si-Le Lin$^{1}$\footnote{E-mail: linsile01@126.com
 This work was financially supported by the funding (funding number: JAT191145, JAT190437) from
the Fujian Education Department.}}
\author{Chun-Lin Zhang$^{2}$}
\author{Si-Yang Hao$^{3}$}
\author{Pei-Yao Hong$^{1}$}

\affiliation {$^{1}$ Jinshan College of Fujian Agriculture and Forestry University, Fuzhou 350001, China
	\\ 		  $^{2}$ Yango University, Fuzhou 350001, China
	\\		  $^{3}$ Fujian Police College, Fuzhou 350001, China}

\begin{abstract}
  We propose a scheme to directly measure the entangled W states with cross-Kerr nonlinearity mediums. This scheme can measure three-photon, four-photon, and even N-photon W states in one step. Only PBSs and cross-Kerr nonlinearity mediums are used in this scheme, which is feasible for experiments. We describe the three-photon W states measurement process and extend it to four-photon and N-photon W states.
\end{abstract}

\pacs {  }

\maketitle

\section{INTRODUCTION}
Quantum entanglement is one of the important resources for quantum information processing and quantum communication. In the field of quantum information, theories based on quantum mechanics have many interesting developments \cite{MBo2001, MBo2002, MBo2003, MBo2021, MBo2022, MBo2023, MBo2024, MBo2025}. The preparation and measurement \cite{MBo2004, MBo2005, MBo2006, MBo2007} of entangled states are important in practical applications \cite{MBo2008, MBo2009}. 

Although the entanglement of binary systems is well studied, the characterization of entanglement in multipartite systems is still rarely studied. Among the multipartite entangled states, the W states is a classic entangled state and can be used in long-distance communications \cite{MBo2010}. The N-qubit W states can be written as $\left | W \right \rangle\ = \frac{1}{\sqrt{n}}\left ( \left | 0...01 \right \rangle\ + \left | 0...10 \right \rangle\ + ... + \left | 1...00 \right \rangle\ \right )$. The schemes for generating the W states \cite{MBo2011,MBo2012,MBo2013} have been proposed. However, the current measurement methods, such as asymmetric W-type measurement \cite{MBo2010} and polarized light measurement \cite{MBo2014}, are complicated. 

Fortunately, the appearance of the cross-Kerr nonlinearity medium made the measurement of polarized photon number easy \cite{MBo2015, MBo2016}. Using cross-Kerr nonlinearity medium and polarization beam splitter (PBS), the entangled state can be measured. Previous work has shown that a quantum nondemolition detectors (QNDs) with cross-Kerr medium and coherent states can be widely used in the construction of phase gates \cite{MBo2017}, the performance of Bell-state analyzers \cite{MBo2018}, entanglement purification and concentration \cite{MBo2012}, etc. The effect of cross-Kerr medium can be described by an interaction Hamiltonian of the form

\begin{equation}
\widehat{H}_{K} = \hbar \chi \widehat{n}_{a} \widehat{n}_{c}
\end{equation}

where $\widehat{H}_{K}$ is the number operator of mode k$\left ( k=a,c \right )$, and $\hbar\chi$ is the nonlinear coupling strength. Then, the photons in mode c will accumulate a phase shift $\theta=\chi$ t proportional to the number of photons in mode a. Such a medium can be used as an optical switch \cite{MBo2016}. The magnitude of the phase $shift \theta$ can be modified by adjusting the coupling length of cross-Kerr nonlinear medium and mode. But the natural cross-Kerr medium has small nonlinearity, the typical dimensionless amplitude is $\theta\approx10^{-18}$ \cite{MBo2019}. This makes the phase shift too small to measure. Fortunately, there are some ways to make the nonlinear amplitude about $10^{-2}$, such as the transparency of electromagnetic induction (EIT) \cite{MBo2020}. With the measured phase shift of the photons in mode c, the number of photons in mode c can be measured. 

In this paper, we present a simple scheme to measure the N-photon entangled W states. We will first exploit the cross-Kerr nonlinearity medium and PBS to construct the measurement of three-photon W states for polarization qubit. Then we generalize this method to measure the four-photon and N-photon W states.

\section{Direct measurement of the three-photon entangled W states}

In order to describe our scheme for W states measurement, we first take the three-photon W states as an example. In this scheme, we use three cross-Kerr nonlinearity meadiums and four PBSs to complete the measurement (Fig. 1). The input (polarized states to be measured) consists of three polarized photons. The polarization of each photon may be horizontal $\left (\left | H  \right \rangle  \right )$ or vertical $\left (\left | V  \right \rangle  \right )$. Then, the W states can be described by the form:

\begin{equation}
\left | \Psi  \right \rangle = \alpha \left | H  \right \rangle \left | H  \right \rangle \left | V  \right \rangle + \beta \left | H  \right \rangle \left | V  \right \rangle \left | H  \right \rangle + \gamma \left | V  \right \rangle \left | H  \right \rangle \left | H  \right \rangle
\end{equation}

The PBS can transmit the horizontally polarized photon $\left | H \right \rangle$, and reflect the vertically polarized photon $\left | V \right \rangle$. We notice that the horizontal symmetrical structure makes the output polarized states equal to the input. In order to measure the number of photons in each mode, we couple a probe beam $\left | \alpha \right \rangle$. with three modes ($a_c$, $b_c$ and $c_c$) by three cross-Kerr nonlinear mediums. When one photon passes through one cross-Kerr nonlinear medium, a phase shift ($+\theta$, $-\theta$ or $+3\theta$) will be introduced to the coherent probe beam $\left | \alpha \right \rangle$. And the probe beam $\left | \alpha \right \rangle$. evolves to $\left | \alpha e^{i\theta} \right \rangle, \left | \alpha e^{-i\theta} \right \rangle or \left | \alpha e^{3i\theta} \right \rangle$. As a result, the total phase shift can be measured by X quadrature homodyne measurement \cite{MBo2018}.

The polarized states of three input photons in modes a, b, and c are $\left|\Psi\right\rangle_a=a_0\left|H\right\rangle_a+a_1\left|V\right\rangle_a$, $\left|\Psi\right\rangle_b=b_0\left|H\right\rangle_b+b_1\left|V\right\rangle_b$ and $\left|\Psi\right\rangle_c=c_0\left|H\right\rangle_c+c_1\left|V\right\rangle_c$. The joint state including three photons will become

\begin{equation}
\left|\left.\Psi\right\rangle_1=(\right.a_0\left|\left.H\right\rangle_a+a_1\left|\left.V\right\rangle_a)\bigotimes{(b}_0\left|\left.H\right\rangle_b+b_1\left|\left.V\right\rangle_b)\bigotimes{(c}_0\left|\left.H\right\rangle_c+c_1\left|\left.V\right\rangle_c)\right.\right.\right.\right.\right.\right.
\end{equation}

When the photons pass through PBS1 and PBS2, the photons are coupled to cross-Kerr nonlinear mediums in modes $a_c$, $b_c$ and $c_c$. The total phase shift of the coherent probe beam $\left|\left.\alpha\right\rangle\right.$ after coupling to three modes ($a_c$, $b_c$ and $c_c$) caused by all possible input polarized states are shown in Table 1.

\begin{table}[]
	\caption{The total phase shift of the coherent probe beam $\left|\left.\alpha\right\rangle\right.$ after coupling to three modes ($a_c$, $b_c$ and $c_c$) caused by all possible input polarized states}
	\begin{tabular}{ccccc}
		\hline
		\multirow{2}{*}{Input} & \multicolumn{3}{l}{The number of photon in modes} & \multirow{2}{*}{Total phase shift $\theta$} \\ 
		& $a_c (+\theta)$& $b_c (-\theta)$& $c_c (+3\theta)$&    \\  \hline 
		
		$\left|H\right\rangle\left|H\right\rangle\left|H\right\rangle$   & 1 & 1  & 1   & 2   \\
		$\left|H\right\rangle\left|H\right\rangle\left|V\right\rangle$   & 1 & 0  & 2   & 7   \\
		$\left|H\right\rangle\left|V\right\rangle\left|H\right\rangle$   & 0 & 2  & 1   & -1  \\
		$\left|H\right\rangle\left|V\right\rangle\left|V\right\rangle$   & 0 & 1  & 2   & 4   \\
		$\left|V\right\rangle\left|H\right\rangle\left|H\right\rangle$   & 2 & 1  & 0   & 0   \\
		$\left|V\right\rangle\left|H\right\rangle\left|V\right\rangle$   & 2 & 0  & 1   & 5   \\
		$\left|V\right\rangle\left|V\right\rangle\left|H\right\rangle$   & 1 & 2  & 0   & -3  \\
		$\left|V\right\rangle\left|V\right\rangle\left|V\right\rangle$   & 1 & 1  & 1   & 2   \\   \hline                                      
	\end{tabular}
\end{table}

The three components of W states $\left( \left|\left.H\right\rangle\right.\left|\left.H\right\rangle\right.\left|\left.V\right\rangle\right.,\ \left|\left.H\right\rangle\right.\left|\left.V\right\rangle\right.\left|\left.H\right\rangle\right.,\ \left|\left.V\right\rangle\right.\left|\left.H\right\rangle\right.\left|\left.H\right\rangle\right.\right)$ can be distinguished from all input polarization states by the total phase shifts of $\left|7\theta\right|$, $\left|-1\theta\right|$ and $\left|0\theta\right|$. Therefore, we can measure the three-photon W states by measuring the phase shift of the coherent probe beam $\left|\left.\alpha\right\rangle\right.$ in only step.

\section{Four-photon and N-photon entangled W states}

Next, we extend the scheme of measuring the three-photon W states to four-photons and N-photon. In the four-photon W-states measurement, we added a PBS3 and a cross-Kerr nonlinear medium in the left side (Fig. 2). Correspondingly, a PBS3’ will be added on the right side. In order to make the input and output polarization states the same, the overall structure is still bilaterally symmetric. Unlike the three-photon W states, the phase shift caused by coupling length of the cross-Kerr nonlinear mediums becomes $+\theta$, $-2\theta$, $+5\theta$ and $-8\theta$. The total phase shift of the coherent probe beam $\left|\left.\alpha\right\rangle\right.$ after coupling to four modes ($a_c$, $b_c$, $c_c$ and $d_c$) caused by all possible input polarized states are shown in Table 2.

\begin{table}[h]
	\caption{The total phase shift of the coherent probe beam $\left|\left.\alpha\right\rangle\right.$ after coupling to four modes ($a_c$, $b_c$, $c_c$ and $d_c$) caused by all possible input polarized states.}
	\begin{tabular}{cccccc}
		\hline
		\multirow{2}{*}{Input} & \multicolumn{4}{c}{The number of photon in modes} & \multirow{2}{*}{Total phase shifts $\theta$} \\
		& $a_c (+\theta)$  & $b_c (-2\theta)$  & $c_c (+5\theta)$  & $d_c (-8\theta)$  &    \\  \hline
		$\left|H\right\rangle\left|H\right\rangle\left|H\right\rangle\left|H\right\rangle$  & 1  & 1  & 1  & 1  & -4 \\
		$\left|H\right\rangle\left|H\right\rangle\left|H\right\rangle\left|V\right\rangle$  & 1  & 1  & 0  & 2  & -17\\
		$\left|H\right\rangle\left|H\right\rangle\left|V\right\rangle\left|H\right\rangle$  & 1  & 0  & 2  & 1  & 3  \\
		$\left|H\right\rangle\left|H\right\rangle\left|V\right\rangle\left|V\right\rangle$  & 1  & 0  & 1  & 2  & -10\\
		$\left|H\right\rangle\left|V\right\rangle\left|H\right\rangle\left|H\right\rangle$  & 0  & 2  & 1  & 1  & -7 \\
		$\left|H\right\rangle\left|V\right\rangle\left|H\right\rangle\left|V\right\rangle$  & 0  & 2  & 0  & 2  & -20\\
		$\left|H\right\rangle\left|V\right\rangle\left|V\right\rangle\left|H\right\rangle$  & 0  & 1  & 2  & 1  & 0  \\
		$\left|H\right\rangle\left|V\right\rangle\left|V\right\rangle\left|V\right\rangle$  & 0  & 1  & 1  & 2  & -13\\
		$\left|V\right\rangle\left|H\right\rangle\left|H\right\rangle\left|H\right\rangle$  & 2  & 1  & 1  & 0  & 5  \\
		$\left|V\right\rangle\left|H\right\rangle\left|H\right\rangle\left|V\right\rangle$  & 2  & 1  & 0  & 1  & -8 \\
		$\left|V\right\rangle\left|H\right\rangle\left|V\right\rangle\left|H\right\rangle$  & 2  & 0  & 2  & 0  & 12 \\
		$\left|V\right\rangle\left|H\right\rangle\left|V\right\rangle\left|V\right\rangle$  & 2  & 0  & 1  & 1  & -1 \\
		$\left|V\right\rangle\left|V\right\rangle\left|H\right\rangle\left|H\right\rangle$  & 1  & 2  & 1  & 0  & 2  \\
		$\left|V\right\rangle\left|V\right\rangle\left|H\right\rangle\left|V\right\rangle$  & 1  & 2  & 0  & 1  & -11\\
		$\left|V\right\rangle\left|V\right\rangle\left|V\right\rangle\left|H\right\rangle$  & 1  & 1  & 2  & 0  & 9  \\
		$\left|V\right\rangle\left|V\right\rangle\left|V\right\rangle\left|V\right\rangle$  & 1  & 1  & 1  & 1  & -4 \\
		\hline
	\end{tabular}
\end{table}

The four components of W states ($\left | H \right \rangle \left | H \right \rangle \left | H \right \rangle \left | V \right \rangle$, $\left | H \right \rangle \left | H \right \rangle \left | V \right \rangle \left | H \right \rangle$, $\left | H \right \rangle \left | V \right \rangle \left | H \right \rangle \left | H \right \rangle$, $\left | V \right \rangle \left | H \right \rangle \left | H \right \rangle \left | H \right \rangle$) can be distinguished from all input polarization states by the total phase shifts of $\left|-17\right|\theta$, $\left|3\right|\theta$, $\left|-7\right|\theta$ and $\left|5\right|\theta$. Therefore, we can measure the four-photon W states by measuring the phase shift of the coherent probe beam $\left|\left.\alpha\right\rangle\right.$ in only one step. In fact, the phase shift of the cross-Kerr nonlinear mediums can be carefully selected to distinguish the W states with a total phase shift of $\left|\left.\alpha\right\rangle\right.$. Based on this scheme, we can extend it to measure the N-photon W states. The schematic diagram is shown in Fig. 3.

With this scheme, we need 2(n-1) PBSs and n well selected cross-Kerr nonlinear mediums to measure N-photon W states. In fact, the key here is to reasonably choose the magnitude of the phase shift of each cross-Kerr medium so that the total phase shift of the target state is distinguished among all possible total phase shifts. With one step measurement of the total phase shift of the probe beam $\left|\left.\alpha\right\rangle\right.$, we can pick out the N-photon W states from all of the input sources.

\section{Discussion and conclusion}

Till now, we have proposed a scheme for measuring N-photon W states. The key step of this scheme is to select the appropriate phase shifts caused by the cross-Kerr nonlinear mediums. We note that since X quadrature homodyne measurement cannot distinguish $+\theta$ and $-\theta$ \cite{MBo2012}, two phase shifts with the absolutely same value cannot be distinguished.. In fact, the phase shift caused by cross-Kerr nonlinear medium is small \cite{MBo2019}, but it is still enough to distinguish different phase shift in the probe beam $\left|\left.\alpha\right\rangle\right.$ \cite{MBo2012}. In the experiment, different phase shifts caused by the cross-Kerr nonlinear mediums can be achieved by increasing or decreasing the coupling length of the cross-Kerr nonlinear medium and the mode.

In summary, we have proposed a scheme to directly measure the N-photon entangled W states with only one step. Only common PBSs and cross-Kerr nonlinear mediums are used in this scheme, which is feasible for future experiments. This scheme may be useful in quantum information processing in the future.

\begin{figure}[h]
	\scalebox{1.0}{\includegraphics {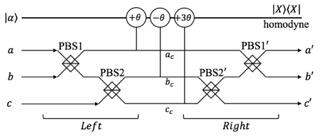}}
	\caption{Schematic diagram of the three-photon entangled W states measurement. The PBS can transmit the horizontally polarized photon $\left| H\right\rangle$, and reflect the vertically polarized photon $\left| V \right \rangle$. A photon in $a_c$ mode introduces $+\theta$ on the coherent probe beam $\left|\alpha\right\rangle$, while a photon in $b_c$ or $c_c$ mode introduces $-\theta$ or $+3\theta$ on $\left|\alpha\right\rangle$, respectively. We make the horizontal structure, Left and Right,  symmetrical to make the polarized state of input and output the same. $\left| X \right \rangle \left\langle X \right|$. represents an X quadrature homodyne measurement. }
	\label{fig1}
\end{figure}
\begin{figure}[h]
	\scalebox{1.0}{\includegraphics {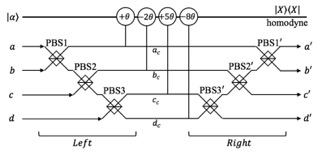}}
	\caption{Schematic diagram of the four-photon entangled W states measurement. A photon in $a_c$ mode introduces $+\theta$ on the coherent probe beam $\left|\left.\alpha\right\rangle\right.$, while a photon in $b_c$, $c_c$ or $d_c$ mode introduces $-2\theta$, $+5\theta$ or $-8\theta$ on $\left|\left.\alpha\right\rangle\right.$, respectively. We make the horizontal structure, Left and Right, symmetrical to make the polarized state of input and output the same. $\left|\left.X\right\rangle\left\langle\left.X\right|\right.\right.$ represents an X quadrature homodyne measurement.}
	\label{fig2}
\end{figure}
\begin{figure}[h]
	\scalebox{1.0}{\includegraphics {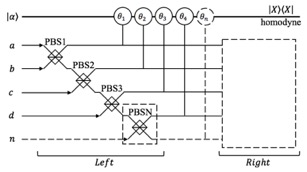}}
	\caption{Schematic diagram of the N-photon entangled W states measurement. If we want to measure the polarization state of one more photon, we can add a mode n and a PBSN to the scheme on the left side. The phase shift of the cross-Kerr nonlinear mediums can be carefully selected to distinguish the W states with a total phase shift of $\left|\left.\alpha\right\rangle\right.$. In order to make the input and output polarizations the save, the overall structure is still bilaterally symmetrical.}
	\label{fig3}
\end{figure}

\section{ACKNOWLEDGMENTS}
  This work was supported by the Young and Middle-aged Teacher Education Research Project of Fujian Province under Grants No. JAT191145, the Sile Lin Project, and Grants No. JAT190437, the Siyang Hao Project.

\end{document}